\documentstyle[aps,preprint]{revtex}

\def\text#1{\mbox{#1}}

\def\ee{\end{equation}}
\def\be#1{\begin{equation}\label{#1}}

\tightenlines
\begin{document}

\title{{Topologically massive nonabelian BF models in arbitrary space-time dimensions}}
\author{ R. R. Landim, C. A. S. Almeida}
\address{Universidade Federal do Cear\'{a}\\
 Physics Department\\
 C.P. 6030, 60470-455
Fortaleza-Ce, Brazil\footnote{ Electronic addresses:
carlos@fisica.ufc.br, renan@fisica.ufc.br}}

\maketitle

\begin{abstract}
This work extends to the D-dimensional space-time the topological
mass generation mechanism of the nonabelian BF model in four
dimensions. In order to construct the gauge invariant nonabelian
kinetic terms for a ($D$-2)-form $B$ and a 1-form $A$, we
introduce an auxiliary ($D$-3)-form $V$. Furthermore, we obtain a
complete set of BRST and anti-BRST transformation rules of the
fields using the so called horizontality condition, and construct
a BRST/anti-BRST invariant quantum action for the model in
$D$-dimensional space-time.

\end{abstract}

\vspace{1.0cm}

PACS: 11.15.-q, 11.10.Ef, 11.10.Kk

\vspace{0.3cm}

Keywords: topological mass generation; nonabelian gauge theories;
antisymmetric tensor gauge fields; arbitrary space-time
dimensions; BRST/anti-BRST symmetry.

\vspace{1.0cm}

Antisymmetric tensor gauge fields appear naturally in string theory and play
an important role in dualization~\cite{ha-si}. They are also fundamental in
realization of Schwarz topological field theory through a BF term, where $B$
is a two form gauge field and $F$ the field strength of the one form gauge
field $A$. The BF term can be abelian or nonabelian, can live in any
dimension, and is a dimensional generalization of the Chern-Simons term.
Some time ago, Allen, Bowick and Lahiri~\cite{La}, using the BF term, showed
that it is possible to give mass to abelian vector gauge fields without the
Higgs field in four dimensions. This interesting mechanism is known as
topological mass generation (TMG). This is a modified form of the well known
topological mass mechanism introduced by Deser and Jackiw in the abelian
gauge models with Chern-Simons term \cite{jackiw}. In the three-dimensional
case a similar mechanism exists for generating mass to the vector field and
alternatively to a scalar and two form field~\cite{non-chern,susy}.
Recently, Hwang and Lee~\cite{H-L}, showed that it is possible to construct
the nonabelian TMG in four dimensions with the addition of an auxiliary
vector field. This auxiliary field is necessary to eliminate a constraint
that appears in the nonabelian version of TMG.

Dimensional generalizations of BF models have been considered in refs. \cite
{lucchesi,baulieu,pisar}. These studies considering strictly Schwarz-type
topological models - which have a classical gauge fixed action written as
the sum of a gauge-invariant term and a BRST-invariant term \cite{blau} -
are focused on perturbative renormalization, symmetry content and formal
aspects of BF models. More recently, Smailagic and Spallucci, have studied
the dualization of abelian \cite{spa} and nonabelian \cite{spa1} BF models
of arbitrary $p-$forms to a Stueckelberg-like massive gauge invariant
theories.

However our main purpose in this letter resides in a slightly different
context. Here we present a $D$-dimensional generalization from the
nonabelian topological mass generation in four dimensions. As mentionated
above this mechanism was introduced by Allen, Bowick and Lahiri for the
abelian BF model in the four dimensional space-time and later applied for
its nonabelian version \cite{H-L}. Using the BRST/anti-BRST formalism we
construct here a framework which consistently prove how the auxiliary fields
are required and unties the constraints in the $D$-dimensional case.

Our starting Lagrangian describes a nonabelian BF model in $D$-dimensions.
Consider the field theory of a real-valued $(D-2)-$form field $B$ and a
real-valued $1-$form field $A$ defined on a D-dimensional space-time
manifold ${\cal M}_D$ with metric $g_{\mu \nu }=\mbox{diag}(-++\cdots +++)$.
Augmented with propagation terms for the gauge fields $A$ and $B$, we have
\begin{equation}
S=\int_{{\cal M}_D}\mbox{Tr}\left( \frac 12H\wedge ^{*}\!\!H+mB\wedge
F-\frac 12F\wedge ^{*}\!\!F\right) ,  \label{act}
\end{equation}
where $F=dA+A\wedge A$ and $H=DB=dB+[A,B]$ are the field strengths of $A$
and $B$ respectively, $d=dx^\mu (\partial /\partial x^\mu )$ is the exterior
derivative and $*$ is the Hodge star operator. Also $A=A^aT^a$, $B=B^aT^a$,
where $T^a$ are generators of a Lie algebra ${\cal G}$ of a semi-simple Lie
group $G$ \footnote{%
The commutator between two Lie algebra valued forms $P$ and $Q$ is defined
by $[P,Q]=P\wedge Q-(-1)^{d(P)d(Q)}Q\wedge P$, where $d(X)$ is the form
degree of $X$.}. Note that $A$ is a $G$-connection $1-$form field on ${\cal M%
}_D$. The adjoint operator acting in a $p-$form can be written as $%
d^{\dagger }=(-1)^{Dp+D}*d*$ (for a Lorentzian manifold) and the Laplacian
operator reads as $\partial ^2=d$ $d^{\dagger }+$ $d^{\dagger }d$ \cite
{nakahara}.

The $B\wedge F$ term in (\ref{act}) is invariant under the gauge
transformations
\begin{eqnarray}
&\delta A=D\theta =d\theta +[A,\theta ],  \nonumber \\
&\delta B=D\Omega +[B,\theta ],  \label{trans}
\end{eqnarray}
where $\theta $ is a $0-$form and $\Omega $ is a $(D-3)-$form Lie algebra
valued. However as in four dimensional case, the action (\ref{act}) is not
invariant under (\ref{trans}), due to the fact that $H$ does not transform
as $\delta H=[H,\theta ]$. In fact, using (\ref{trans}), the transformation
of $H$ is
\begin{equation}
\delta H=[F,\Omega ]+[H,\theta ].  \label{Htr}
\end{equation}
In order to circumvent this problem, at least classically, we introduce an
auxiliary $(D-3)-$form $V$, and redefine $H$ to be
\begin{equation}
{\cal H}=H+[F,V],  \label{newH}
\end{equation}
with $V$ transforming as
\begin{equation}
\delta V=-\Omega +[V,\theta ].  \label{Vtran}
\end{equation}

It is important to remark that for any dimension of space-time, the
covariant derivative must be constructed with a $1-$form $A$ \cite{thierry}.

This procedure is a generalization of the mechanism first introduced by
Thierry-Mieg {\it et al. }\cite{thierry,thierry-baulieu} who detected the
obstruction to the nonabelianization of the term $H\wedge ^{*}\!\!H$. As
pointed out by Hwang and Lee \cite{H-L} in four dimensional case, the
equation of motion given by the action (\ref{act}), namely $D^{*}\!\!H+mF=0,$
gives the constraint

\begin{equation}
DD^{*}\!\!H=[F,^{*}\!\!H]=0  \label{const}
\end{equation}
which is solved when we introduce the auxiliary $(D-3)-$form $V$.

To implement the quantization of the model, we have to construct the BRST
and anti-BRST symmetry .

In the work of Thierry-Mieg and Ne'eman \cite{thierry}, a geometrical BRST
quantization scheme was developed where the base space is extended to a
fiber bundle space so that it contains unphysical (fiber-gauge orbit)
directions and physical (space-time) directions. Using a double fiber bundle
structure Quiros {\it et al. }\cite{quiros} extended the principal fiber
bundle formalism in order to include anti-BRST symmetry. Basically the
procedure consists in extending the space-time to take into account a pair
of scalar anticommuting coordinates denoted by $y$ and $\overline{y}$ which
correspond to coordinates in the directions of the gauge group of the
principal fiber bundle. Then the so-called ''horizontality condition'' is
imposed. This condition enforces the curvature components containing
vertical (fiber) directions to vanish. Hence only the horizontal components
of physical curvature in the extended space survive.

Let us define the following form fields in the extended space and valued in
the Lie algebra ${\cal G}$ of the gauge group:
\begin{eqnarray}
\widetilde{A} &=A+c+\bar{c},  \label{e1} \\
\widetilde{F} &=\widetilde{d}\widetilde{A}+\widetilde{A}\wedge \tilde{A},
\label{e2} \\
\widetilde{B} &=\sum_{k=0}^{D-2}\sum_{n=0}^kB_{D-2-k}^{(n)(k-n)},
\label{e3} \\
\widetilde{{\cal H}} &=\widetilde{D}\widetilde{B}+\left[ \widetilde{F},%
\widetilde{V}\right] ,  \label{e4} \\
\widetilde{V} &=\sum_{k=0}^{D-3}\sum_{n=0}^kV_{D-3-k}^{(n)(k-n)},
\label{e5} \\
\widetilde{D} &=\widetilde{d}+[\widetilde{A},~~]~,  \label{e6} \\
\widetilde{d} &=d+s+\bar{s}.  \label{e7}
\end{eqnarray}

Here we identify the components in unphysical directions with new fields,
namely, $c$ ($\overline{c}$) as ghosts (antighosts) in the case of the field
$A.$ There are $D(D-1)/2$ and $(D-1)/(D-2)/2$ components in $\widetilde{B}$
and $\widetilde{V}$ respectively. In the expansion of $\widetilde{B}$ and $%
\widetilde{V}$, the upper indices $n$ and $k-n$ are respectively the ghost
number and the antighost number of the $(D-2-k)-$form $B_{D-2-k}^{(n)(k-n)}$
and $(D-3-k)-$form $V_{D-3-k}^{(n)(k-n)}$, having a total degree $(D-2)$ and
$(D-3)$.

Note that when we treat two odd ''extended'' forms, the $[$ $,$ $]$ must be
reading as an anticommutator.

Furthermore, we call attention for the necessary presence of the auxiliary $%
(D-3)-$form $V$ field. The exterior derivatives in the gauge group
directions are denoted by $s=dy^N(\partial /\partial y^N)$ and $\overline{s}%
=d\overline{y}^{\overline{N}}(\partial /\partial \overline{y}^{\overline{N}%
}).$

The horizontality condition, or equivalently, the Maurer-Cartan equation for
the field strengths $F$ and ${\cal H}$ can be written as
\begin{eqnarray}
\widetilde{{\cal H}} &={\cal H},  \label{HT} \\
\widetilde{F} &=F.  \label{FT}
\end{eqnarray}
The expansion of (\ref{FT}), gives us the well known BRST and anti-BRST of
the gauge field $A$, the ghost $c$ and the anti-ghost $\bar{c}$:
\begin{eqnarray}
sA &=-Dc,\quad \bar{s}A=-D\bar{c}  \nonumber \\
sc &=-cc,\quad \bar{s}\bar{c}=-\bar{c}\bar{c}~~~  \label{TA} \\
s\bar{c}&+\bar{s}c =-[c,\bar{c}]~~~~~  \nonumber
\end{eqnarray}
The transformation BRST and anti-BRST in the last equation of (\ref{TA}),
are unknown. We must introduce an auxiliary field $b$, in order to fix
completely those transformations
\begin{equation}
s\bar{c}=b,\quad \bar{s}c=-b-[c,\bar{c}],\quad sb=0,\quad \bar{s}b=-[\bar{c}%
,b]  \label{bt}
\end{equation}
Now, expanding the equation (\ref{HT}) into a basis of same ghost number and
form degree, we have
\begin{equation}
sB_{D-2-k}^{(k)(0)}=-DB_{D-3-k}^{(k+1)(0)}-[c,B_{D-2-k}^{(k)(0)}]-[F,V_{D-4-k}^{(k+1)(0)}],
\label{BRST-Bk}
\end{equation}
\begin{equation}
\bar{s}B_{D-2-k}^{(0)(k)}=-DB_{D-3-k}^{(0)(k+1)}-[\bar{c}%
,B_{D-2-k}^{(0)(k)}]-[F,V_{D-4-k}^{(0)(k+1)}],  \label{a-BRSTBK}
\end{equation}
for $0\le k\le D-2$, and
\begin{eqnarray}
DB_{D-3-k}^{(n)(k-n+1)}+[c,B_{D-2-k}^{(n-1)(k-n+1)}]+[\bar{c}%
,B_{D-2-k}^{(n)(k-n)}]+  \nonumber \\
sB_{D-2-k}^{(n-1)(k-n+1)}+\bar{s}%
B_{D-2-k}^{(n)(k-n)}+[F,V_{D-4-k}^{(n)(k-n+1)}] =0,  \label{TB1}
\end{eqnarray}
for $1\le n\le k$, $1\le k\le D-2$.

We have $(D-1)(D-2)$ not defined $s$ and $\bar{s}$ transformations in (\ref
{TB1}). Therefore we must introduce a set of $(D-1)(D-2)/2$ auxiliary
fields, namely $\omega ^{\prime }s,$ in order to fix the transformation rule
completely.
\begin{equation}
sB_{D-2-k}^{(n-1)(k-n+1)}=\omega _{D-2-k}^{(n)(k-n+1)}  \label{SB}
\end{equation}
\begin{eqnarray}
\bar{s}B_{D-2-k}^{(n)(k-n)} &=-DB_{D-3-k}^{(n)(k-n+1)}-\omega
_{D-2-k}^{(n)(k-n+1)}-~~~~  \nonumber \\
&[c,B_{D-2-k}^{(n-1)(k-n+1)}]-[\bar{c}%
,B_{D-2-k}^{(n)(k-n)}]-[F,V_{D-4-k}^{(n)(k-n+1)}]  \label{ASB}
\end{eqnarray}

The condition (\ref{HT}) leads us to
\begin{equation}
\widetilde{B}+\widetilde{D}\widetilde{V}=B+DV,  \label{BDV}
\end{equation}
which now yields the BRST/anti-BRST transformation rule for the components
of $\widetilde{V}$
\begin{equation}
sV_{D-3-k}^{(k)(0)}=-DV_{D-4-k}^{(k+1)(0)}-[c,V_{D-3-k}^{(k)(0)}]-B_{D-3-k}^{(k+1)(0)},
\label{SVK}
\end{equation}
\begin{equation}
\bar{s}V_{D-3-k}^{(0)(k)}=-DV_{D-4-k}^{(0)(k+1)}-[\bar{c}%
,V_{D-3-k}^{(0)(k)}]-B_{D-3-k}^{(0)(k+1)},  \label{ASVK}
\end{equation}
for $0\le k\le D-3$, and
\begin{eqnarray}
DV_{D-4-k}^{(n)(k-n+1)}+sV_{D-3-k}^{(n-1)(k-n+1)}+\bar{s}%
V_{D-3-k}^{(n)(k-n)}+~~~~~  \nonumber \\
~[c,V_{D-3-k}^{(n-1)(k-n+1)}]+[\bar{c}%
,V_{D-3-k}^{(n)(k-n)}]+B_{D-3-k}^{(n)(k-n+1)} =0,
\end{eqnarray}
for $1\le n\le k$ and $1\le k\le D-3$. Again, these equations do not fix the
BRST/anti-BRST transformation rule, so we need a set of $(D-2)(D-3)/2$
auxiliary fields $\eta $:
\begin{equation}
sV_{D-3-k}^{(n-1)(k-n+1)}=\eta _{D-3-k}^{(n)(k-n+1)}  \label{SV}
\end{equation}
\begin{eqnarray}
\bar{s}V_{D-3-k}^{(n)(k-n)} &=-\eta
_{D-3-k}^{(n)(k-n+1)}-DV_{D-4-k}^{(n)(k-n+1)}-  \nonumber \\
&[c,V^{(n-1)(k-n+1)}]-[\bar{c}%
,V_{D-3-k}^{(n)(k-n)}]-B_{D-3-k}^{(n)(k-n+1)}.
\end{eqnarray}
In order to obtain the BRST/anti-BRST transformations of the auxiliary
fields $\omega $ and $\eta $, we use the nilpotency condition of $s$ and $%
\bar{s}$:
\begin{equation}
s\omega _{D-2-k}^{(n)(k-n+1)}=0  \label{somega}
\end{equation}
\begin{eqnarray}
\bar{s}\omega _{D-2-k}^{(n)(k-n+1)}&=-[Dc,B_{D-3-k}^{(n-1)(k-n-2)}]-D\omega
_{D-3-k}^{(n)(k-n+2)}-  \nonumber \\
&[B_{D-2-k}^{(n-1)(k-n+1)},b]-[\bar{c},\omega
_{D-2-k}^{(n)(k-n+1)}]-[c,\omega _{D-2-k}^{(n-1)(k-n+2)}]- \\
&[cc,B_{D-2-k}^{(n-2)(k-n+2)}]+[F,\eta _{D-4-k}^{(n)(k-n+2)}]  \nonumber
\end{eqnarray}
\begin{equation}
s\eta _{D-3-k}^{(n)(k-n+1)}=0  \label{seta}
\end{equation}
\begin{eqnarray}
\bar{s}\eta _{D-3-k}^{(n)(k-n+1)}&=-[Dc,V_{D-4-k}^{(n-1)(k-n+2)}]-D\eta
_{D-3-k}^{(n)(k-n+1)}+  \nonumber \\
&\omega _{D-3-k}^{(n)(k-n+2)}-[c,\eta _{D-3-k}^{(n-1)(k-n+2)}]-[\bar{c}%
,\eta _{D-3-k}^{(n)(k-n+1)}]-  \label{seta1} \\
&[cc,V_{D-3-k}^{(n-2)(k-n+2)}]-[V_{D-3-k}^{(n-1)(k-n+1)},b]  \nonumber
\end{eqnarray}

Therefore, a complete set of BRST and anti-BRST equations, namely, eqs. (\ref
{TA}-\ref{a-BRSTBK}), (\ref{SB},\ref{ASB}), (\ref{SVK}-\ref{ASVK}), and (\ref
{SV}-\ref{seta1}), associated with the classical symmetry (\ref{trans}),
were obtained.

Finally, the topologically massive nonabelian BF model in $D-$dimensions
BRST/anti-BRST invariant can be written as
\begin{equation}
S_{cl}=\int_{{\cal M}_D}\mbox{Tr}\left( \frac 12{\cal H}\wedge ^{*}\!\!{\cal %
H}+mB\wedge F-\frac 12F\wedge ^{*}\!\!F\right) .  \label{act1}
\end{equation}
In the expression above we could be use ${\cal B}=B+DV.$ However, in the
action (\ref{act1}), $mB\wedge F$ differs from $m{\cal B}\wedge F$ by a
total derivative term as occurs in the four dimensional case.

The simplest scenario to study mass generation is to consider the equations
of motion of the action (\ref{act1}). Namely
\begin{equation}
(-1)^{(D-1)}D^{*}{\cal H}+mF=0  \label{3.1}
\end{equation}
and
\begin{equation}
D^{*}F=mDB+[B,^{*}{\cal H]}+D[V,^{*}{\cal H}].  \label{3.2}
\end{equation}

\begin{equation}
\lbrack F,^{*}{\cal H}]=0  \label{3.3}
\end{equation}
The last equation corresponds to the constraint (\ref{const}), but now
appears as the equation of motion for the auxiliary field $V.$

Considering only linear terms for the fields, from equations (\ref{3.1}) and
(\ref{3.2}) we get:
\begin{equation}
\left( \partial ^2-m^2\right) H=0,  \label{3.4}
\end{equation}
\begin{equation}
\left( \partial ^2-m^2\right) F=0.  \label{3.5}
\end{equation}
which exhibit mass generation for $H$ and $F.$ Note that, in our metric, $%
\partial ^2=-\partial _t^2+\nabla ^2=-\Box $.

We now propose as a quantized action of our $D-$dimensional nonabelian BF
model the following expression

\begin{equation}
S=S_{cl}+\int \mbox{Tr}s\overline{s}\left( A\wedge ^{*}A+\alpha c\wedge ^{*}%
\overline{c}+\sum_{k=0}^{D-2}\sum_{n=0}^{k-1}\lambda
_{kn}B_{D-2-k}^{(n)(k-n)}\wedge ^{*}B_{D-2-k}^{(k-n)(n)}\right) ,
\label{qaction}
\end{equation}
where $\alpha $ and $\lambda _{kn}$ are gauge parameters.

This geometrical quantization method was formulated by
Thierry-Mieg and Baulieu \cite{thierry-baulieu2} for Yang-Mills
theories and later for nonabelian antisymmetric tensor gauge
theories \cite{thierry-baulieu}. This method is particularly
relevant for treating models with ghosts for ghosts and
differential constraints, where the Fadeev-Popov construction does
not work.

It is worth mentioning that years ago, a nonabelian dimensional
generalization of topologically massive gauge theories involving
antisymmetric tensor fields was proposed \cite{oda}. However it {\it does
not describe} a BF $D-$dimensional model. In fact, that work consider $n-$%
form and ($D-n-1$) form fields (which are not gauge connections) and does
not present an Yang-Mills term $TrF^2$ . As a matter of fact, they consider
a flat connection 1-form $A$, consequently $F=0.$ Therefore the constraints
above discussed are absent from that model.

In summary, we setup a geometrical construction of the BRST/anti-BRST
equations in the massive gauge-invariant nonabelian BF model in $D$ $-$
dimensions using the horizontality condition. This procedure generalizes the
method of Hwang and Lee in four dimensional space-time \cite{H-L}, and
consistently extend to $D-$ dimensions the introduction of auxiliary fields
in order to provide a nonabelianization of the propagation term of a $(D-2)-$%
form field.

\end{document}